\documentclass[12pt]{article}

\usepackage{amsmath}

\begin{document}

\begin{center}
{\bf
On conserved quantities in the  theory 
 of charged boson fields of spin 0 and 1
     \\ [3mm]
    Tokarevskaya N.G., Red'kov V.M.\footnote{   E-mail:
redkov@dragon.bas-net.by }\\[3mm]}
Institute of Physics, National  Academy of Sciences
of Belarus \\
68 Skarina  str.,  Minsk,  BELARUS

\end{center}

\begin{quotation}
Bosons of spin 0 and 1,  with different intrinsic parities, are
described by full sets of spinor equations in  the frame of the
Dirac-Kahler theory. This enables us to obtain the conservation
laws  for the boson particles with one value of spin by imposing
linear additional  conditions in the known sixteen conserved
currents of the Dirac-Kahler field. In this way for each boson the
known conserved quantities, charge vector $j^{a}(x)$, symmetrical
energy-momentum ten\-sor $T^{ab}(x)$, and  angular moment ten\-sor
$L^{a[bc]}(x)$,  have been found. Additionally, for scalar fields,
one  conserved current $\nu^{a}(x)$ has been constructed; it is
not zero one only for a complex-valued field. For a vector
particles, two additional currents,  $\nu^{a}(x)$ and
$\nu^{ab}(x)$, are  found  that again do not vanish  when fields
are complex-valued. Those currents  $\nu^{a}(x)$ and $\nu^{ab}(x)$
have not  seemingly any physical interpretation.

\end{quotation}

\section{Introduction }

It is known that the wave  equation of the Dirac-Kahler  field
allows for sixteen conservation laws and there exists a number of
works where such conserved Dirac-Kahler currents have been
examined (for example, see in [1-19]). However previously, as a
rule, the main attention was given to existence  of those
conservation laws and description of its underlying symmetry
instead of elucidating its physical meaning. Else one point
significant for us is that  the Dirac-Kahher theory gives
possibility to obtain many facts of  the more simple theory of
ordinary boson particles of spin 0 or 1 ;  this is achieved by
imposing special  linear conditions on 16-component
 Dirac-Kahler wave function [20]. The same trick may be done
with the  Dirac-Kahler conserved currents, ant it is the  object
of the present note.

\section{Conserved quantities in the theory of the
Dirac-Kahler  field}

\hspace{5mm} Before proceeding to the problem of conserved
quantities in the Dirac-Kahler  theory, with subsequent analysis
of the most interest cases of ordinary bosons particles of spin 0
or 1 and different intrinsic parity,  let us consider in the
beginning the simple example of the  Dirac  particle -- at this
there exists possibility ro introduce the notation required.

 The Dirac equation and its conjugate for $\bar{\Psi}(x) = \Psi^{+}(x)  \gamma^{0} $
have the form \begin{eqnarray}
 [\; i\gamma^{a}
\stackrel{\rightarrow}{\partial}_{a} - m \; ]\; \Psi (x) = 0 \; ,
\qquad \bar{\Psi}(x) \; [\; i\gamma^{a}
\stackrel{\leftarrow}{\partial}_{a} + m\; ] = 0 \; .
\label{2.1}
\end{eqnarray}

\noindent In the following we are to take the formulas (the use of
spinor basis is presupposed)
\begin{eqnarray}
(\gamma^{0})^{+} =  +\; \gamma^{0}, \qquad (\gamma^{i})^{+} = -\;
\gamma^{i} \; , \qquad \gamma^{0} (\gamma^{a})^{+} \gamma^{0} = +
\gamma^{a} \; .
\label{2.2}
\end{eqnarray}

\noindent From (\ref{2.1}) through  the  known procedure it follows the
charge conservation law
\begin{eqnarray}
\partial_{a}  J^{a}(x) = 0 , \qquad  \mbox{где} \qquad J^{a}(x) = \bar{\Psi}(x) \gamma^{a} \Psi (x) \; .
\label{2.3}
\end{eqnarray}

The symmetry properties of the  vector $J^{a}(x)$  under double
spin covering of the full Lorentz group $G_{L}^{spin}$
$$
S(k,\bar{k}^{*}) \; , \qquad  M = i  \gamma^{0} \;, \qquad  N =
\gamma^{0} \gamma^{5}
$$

\noindent can be easily described:
\begin{eqnarray}
\left. \begin{array}{ll} S(k,\bar{k}^{*}): & \qquad
J\; ^{'a}(x') = L^{a}_{\;\;\;\;b}(k,k^{*})  \; J^{b}(x) \; , \\[2mm]
M : & \qquad
J\; ^{'a}(x') = +\; L^{(P)a}_{\;\;\;\;\;\;\;b} \; J^{b}(x) \; , \\[2mm]
N : & \qquad J\; ^{'a}(x') = - \; L^{(T)a}_{\;\;\;\;\;\;\;\;b} \;
J^{b}(x) \; .
\end{array} \right.
\label{2.7}
\end{eqnarray}

The  problem of conserved  currents in the Dirac-Kahler theory may
be considered in much the same  manner. The main equation and its
conjugate are
\begin{eqnarray}
[\; i \gamma^{a} \stackrel{\rightarrow}{\partial}_{a} - m \; ]\;
U(x) =0 \; ,  \qquad \bar{U}(x) \; [\; i \gamma^{a}
\stackrel{\leftarrow}{\partial}_{a} + m \; ] =0 \; ,
\label{2.8a}
\\
\bar{U}(x)= U^{+}(x) \; (\gamma^{0} \otimes \gamma^{0} ) =
 \tilde{\gamma}^{0} U^{+}(x) \gamma^{0} \; .
\label{2.8b}
\end{eqnarray}

\noindent The Dirac-Kahler  functions behave under the  group
$G_{L}^{spin}$ as follows:
\begin{eqnarray}
\left. \begin{array}{ll}
S(k,\bar{k}^{*})\; : & \qquad U'(x') = (S \otimes S) \;U(x) = S \;U(x) \;\tilde{S} \; , \\[3mm]
& \qquad \bar{U}'(x') = \bar{U}(x) \; (S^{-1} \otimes S^{-1}) =
\tilde{S}^{-1} \;  \bar{U}(x) \; S^{-1} \; ,
\end{array} \right.
\label{2.9a}
\\[3mm]
\left. \begin{array}{ll}
M = i \gamma^{0} \; : & \qquad U'(x') = (M \otimes M) \;U(x) = M \;U(x) \;\tilde{M} \; , \\[3mm]
& \qquad \bar{U}'(x') = \bar{U}(x) \; (M^{-1} \otimes M^{-1}) =
\tilde{M}^{-1} \;  \bar{U}(x) \; M^{-1} \; ,
\end{array} \right.
\label{2.9b}
\end{eqnarray}

\noindent and
\begin{eqnarray}
\left. \begin{array}{ll} N = \gamma^{0} \gamma^{5} \; : & \qquad
U'(x') = (N \otimes N)
\;U(x) = N \;U(x) \;\tilde{N} \; , \\[3mm]
& \qquad \bar{U}'(x') = \bar{U}(x) \; (N^{-1} \otimes N^{-1}) =
\tilde{N}^{-1} \;  \bar{U}(x) \; N^{-1} \; .
\end{array} \right.
\label{2.9c}
\end{eqnarray}

Combining equations from   (\ref{2.8a}),  we get to sixteen conserved
currents:
\begin{eqnarray}
\bar{U} [\; i \gamma^{a} \stackrel{\rightarrow}{\partial}_{a} - m
\; ]\; U(x) + \bar{U}(x) \; [\; i \gamma^{a}
\stackrel{\leftarrow}{\partial}_{a} + m \; ] U(x) =0 \; :
 \qquad \Longrightarrow
\nonumber
\\
\partial_{a} \; J^{a}(x) =0 \; , \qquad \mbox{where} \qquad
J^{a}(x) = \bar{U} (x)\gamma^{a}  U(x) \; .
\label{2.10a}
\end{eqnarray}

Take notice that  $J^{a}(x)$ is a  $4\times 4$ matrix; that is it
represents 16 conserved quantities.  On the base of these one  may
construct special linear combinations which will be of simple
symmetry properties under the group $G_{L}^{spin}$:
\begin{eqnarray}
\left. \begin{array}{rl}
j^{(a)}(x) = \mbox{Sp} \;J^{a}(x) \; , & \qquad \partial_{a} j^{(a)}(x) = 0   \; ;\\[3mm]
\nu^{(a)}(x) = \mbox{Sp} \;\gamma^{5} \; J^{a}(x) \; ,
& \qquad \partial_{a} \nu^{(a)}(x) = 0   \; ; \\[3mm]
T^{(a)l}(x) = \mbox{Sp} \; \tilde{\gamma}^{l} \; J^{a}(x) \; ,
& \qquad \partial_{a} T^{(a)l}(x) = 0   \; ;\\[3mm]
\nu^{(a)l}(x) = \mbox{Sp} \;\gamma^{5} \tilde{\gamma}^{l}\; \;
J^{a}(x) \; ,
& \qquad \partial_{a} \nu^{(a)l}(x) = 0   \; ;\\[3mm]
L^{(a)kl}(x) = \mbox{Sp} \; \tilde{\sigma}^{kl} \; J^{a}(x) \; , &
\qquad \partial_{a} \varphi^{(a)kl}(x) = 0   \; .
\end{array} \right.
\label{2.10b}
\end{eqnarray}

Now the  task is to follow the symmetry properties of these
separate currents under the  group $G_{L}^{spin}$. First, let us
consider the  current $j^{(a)}(x)$:
\begin{eqnarray}
S(k,\bar{k}^{*}):  \qquad j^{('a)}=
L^{a}_{\;\;\;\;b}(k,\; ,k^{*}) \;
j^{(b)}\; ,
\nonumber
\\
 M:  \qquad j^{('a)}=  L^{(P)a}_{\;\;\;\;\;\;\;b} \;
j^{(b)}\; ,
\nonumber
\\
N:  \qquad j^{('a)}=  = L^{(T)a}_{\;\;\;\;\;\;\;b} \;
j^{(b)}\; .
\label{2.11c}
\end{eqnarray}

For the current  $\nu^{(a)}(x)$
\begin{eqnarray}
S(k,\bar{k}^{*}):  \qquad \nu^{('a)}(x')= + L^{a}_{\;\;\;\;b}(k,\;
,k^{*}) \; \nu^{(b)}(x)\; ,
\nonumber
\\
M:  \qquad \nu^{('a)}(x') =-\; L^{(P)a}_{\;\;\;\;\;\;\;b} \;
\nu^{(b)}(x)\; ,
\nonumber
\\
N:  \qquad \nu^{('a)}(x')=
 -\; L^{(T)a}_{\;\;\;\;\;\;\;b} \;
\nu^{(b)}(x)\; .
\label{2.12c}
\end{eqnarray}

For the current  $T^{(a)l}(x)$
\begin{eqnarray}
S(k,\bar{k}^{*}):  \qquad T^{('a)l}(x')= L^{l}_{\;\;\;k}
\;L^{a}_{\;\;\;\;b} \; T^{(b)k}(x)\; ,
\nonumber
\\
M:  \qquad T^{('a)l}(x')= L^{(P)l}_{\;\;\;\;\;\;\;k}\;
L^{(P)a}_{\;\;\;\;\;\;\;b} \; T^{(b)k}(x)\; ,
\nonumber
\\
N: \qquad T^{('a)l}(x')= L^{(T)l}_{\;\;\;\;\;\;\;k}\;
L^{(T)a}_{\;\;\;\;\;\;\;b} \; T^{(b)k}(x)\; .
\label{2.13c}
\end{eqnarray}

For the current   $\nu^{(a)l}(x)$
\begin{eqnarray}
S(k,\bar{k}^{*}):  \qquad T^{('a)l}(x')= L^{l}_{\;\;\;k}
\;L^{a}_{\;\;\;\;b} \; \nu^{(b)k}(x)\; ,
\nonumber
\\
M:  \qquad \nu^{('a)l}(x')= -\; L^{(P)l}_{\;\;\;\;\;\;\;k}\;
L^{(P)a}_{\;\;\;\;\;\;\;b} \; \nu^{(b)k}(x)\; ,
\nonumber
\\
N:  \qquad \nu^{('a)l}(x')= -\; L^{(T)l}_{\;\;\;\;\;\;\;k}\;
L^{(T)a}_{\;\;\;\;\;\;\;b} \; \nu^{(b)k}(x)\; .
\label{2.14c}
\end{eqnarray}

And  for the current $ L^{'(a)bc}(x')$ we have
\begin{eqnarray}
S(k,\bar{k}^{*}): \qquad L^{'(a)bc}(x')  = L^{a}_{\;\;\;b}
L^{b}_{\;\;\;k} L^{c}_{\;\;\;l} \; L^{(b)kl}(x) \; ,
\nonumber
\\
M: \qquad L^{'(a)bc}(x')  = L^{(P)a}_{\;\;\;\;\;\;\;\;b}
L^{(P)b}_{\;\;\;\;\;\;\;\;k} L^{(P)c}_{\;\;\;\;\;\;\;\;l} \; \;
L^{(b)kl}(x) \; ,
\nonumber
\\
N: \qquad L^{'(a)bc}(x')  =
 L^{(T)a}_{\;\;\;\;\;\;\;\;b}
L^{(T)b}_{\;\;\;\;\;\;\;\;k} L^{(T)c}_{\;\;\;\;\;\;\;\;l} \; \;
L^{(b)kl}(x) \; .
\label{2.15c}
\end{eqnarray}

\section{ Conserved currents for boson fields in tensor form }

\hspace{5mm} Now the  task is to obtain explicit tensor form  for
the above currents (\ref{2.10b}). With the use of the formula for the
bispinor   metric  matrix
\begin{eqnarray}
E = \left | \begin{array}{cc} i\sigma^{2} & 0 \\ 0 & -i\sigma^{2}
\end{array} \right |
 = i\;\gamma^{0}\gamma^{2} \; , \qquad E^{-1} = -\;  i\;\gamma^{0}\gamma^{2} \; ,
\nonumber
\end{eqnarray}

\noindent the Dirac-Kahler function looks as
\begin{eqnarray}
U(x) = \left [\;  - i \; A  +  \gamma ^{5} \; B  +   \gamma^{l} \;
A _{l}  +
           i \; \gamma ^{l} \gamma ^{5} \; B_{l}
+
           i \; \sigma^{mn}\;  F _{mn}   \; \right ]\;(-\;  i\;\gamma^{0}\gamma^{2})  \; ,
\label{3.1a}
\end{eqnarray}

\noindent also
$$
U^{+}(x) = ( i\;\gamma^{0}\gamma^{2})\; \left [ \;
 + i \; A ^{*} +  \gamma ^{5} \; B^{*}  +   (\gamma^{l})^{+} \; A^{*} _{l}  -
           i \; \gamma ^{5} (\gamma ^{l})^{+}  \; B_{l}^{*}
-           i \; (\sigma^{mn})^{+} \;  F ^{*}_{mn}   \; \right ]
\; ,
$$

\noindent and further
\begin{eqnarray}
\bar{U}(x) =
 ( -i\;\gamma^{0}\gamma^{2}) \left [
 + i \; A ^{*} -  \gamma ^{5} \; B^{*}  +   \gamma^{l} \; A^{*} _{l}  +
           i \; \gamma ^{5} \gamma ^{l}  \; B_{l}^{*}
+           i \; \sigma^{mn} \;  F ^{*}_{mn}    \right ] .
\label{3.1b}
\end{eqnarray}

\noindent Therefore, the conserved current matrix may be given as
\begin{eqnarray}
J^{(a)}(x) = \bar{U}(x) \gamma^{a} U(x) =
\nonumber
\\
= ( -i\;\gamma^{0}\gamma^{2})\; \left ( \;
 + i \; A ^{*} -  \gamma ^{5} \; B^{*}  +   \gamma^{p} \; A^{*} _{p}  +
           i \; \gamma ^{5} \gamma ^{p}  \; B_{p}^{*}
+           i \; \sigma^{bc} \;  F ^{*}_{bc}   \; \right )\;
\times
\nonumber
\\
\times \gamma^{a}\; \left (\;  - i \; A  +  \gamma ^{5} \; B  +
\gamma^{s} \; A _{s}  +
           i \; \gamma ^{s} \gamma ^{5} \; B_{s}
+
           i \; \sigma^{mn}\;  F _{mn}   \; \right ) \;(-\;  i\;\gamma^{0}\gamma^{2}) \; .
\label{3.2}
\end{eqnarray}

\noindent Also we  need two relations:
\begin{eqnarray}
\tilde{\gamma}^{l} \; (- \;i\gamma^{0} \gamma^{2}) =  -\;
\gamma^{l} \; ( -i\gamma^{0}\gamma^{2}) \; , \qquad
\tilde{\sigma}^{ab} \;(- \;i\gamma^{0} \gamma^{2})  = -\; (-\;
i\gamma^{0} \gamma^{2}) \; \sigma^{ab} \; . \nonumber
\end{eqnarray}

\noindent So that, from  (\ref{2.10b}) it follows the formulas for
conserved currents of the
 Dirac-Kahler field:
\begin{eqnarray}
j^{(a)}(x) = \mbox{Sp} \;J^{a}(x) =
\nonumber
\\
= -\; \mbox{Sp} \;  \; \left [ \left ( \;
 + i \; A ^{*} -  \gamma ^{5} \; B^{*}  +   \gamma^{p} \; A^{*} _{p}  +
           i \; \gamma ^{5} \gamma ^{p}  \; B_{p}^{*}
+           i \; \sigma^{bc} \;  F ^{*}_{bc}   \; \right )\;
\times \right.
\nonumber
\\
\left. \times \gamma^{a}\; \left (\;  - i \; A  +  \gamma ^{5} \;
B  +   \gamma^{s} \; A _{s}  +
           i \; \gamma ^{s} \gamma ^{5} \; B_{s}
+ i \; \sigma^{mn}\;  F _{mn}   \; \right ) \; \right ] \; ,
\label{3.4a}
\end{eqnarray}
\begin{eqnarray}
\nu^{(a)}(x) = \mbox{Sp} \;\gamma^{5} \; J^{a}(x) = \nonumber
\\
= -\; \mbox{Sp} \;  \; \left [  \gamma^{5} \; \left ( \;
 + i \; A ^{*} -  \gamma ^{5} \; B^{*}  +   \gamma^{p} \; A^{*} _{p}  +
           i \; \gamma ^{5} \gamma ^{p}  \; B_{p}^{*}
+           i \; \sigma^{bc} \;  F ^{*}_{bc}   \; \right )\;
\times \right.
\nonumber
\\
\left. \times \gamma^{a}\; \left (\;  - i \; A  +  \gamma ^{5} \;
B  +   \gamma^{s} \; A _{s}  +
           i \; \gamma ^{s} \gamma ^{5} \; B_{s}
+ i \; \sigma^{mn}\;  F _{mn}   \; \right ) \; \right ] \; ,
\label{3.4b}
\end{eqnarray}
\begin{eqnarray}
T^{(a)l}(x) = \mbox{Sp} \; \tilde{\gamma}^{l} \; J^{a}(x) =
\nonumber
\\
= +\; \mbox{Sp} \;  \; \left [  \gamma^{l}\; \left ( \;
 + i \; A ^{*} -  \gamma ^{5} \; B^{*}  +   \gamma^{p} \; A^{*} _{p}  +
           i \; \gamma ^{5} \gamma ^{p}  \; B_{p}^{*}
+           i \; \sigma^{bc} \;  F ^{*}_{bc}   \; \right )\;
\times \right.
\nonumber
\\
\left. \times \gamma^{a}\; \left (\;  - i \; A  +  \gamma ^{5} \;
B  +   \gamma^{s} \; A _{s}  +
           i \; \gamma ^{s} \gamma ^{5} \; B_{s}
+ i \; \sigma^{mn}\;  F _{mn}   \; \right ) \; \right ] \; ,
\label{3.4c}
\end{eqnarray}
\begin{eqnarray}
\nu^{(a)l}(x) = \mbox{Sp} \;\gamma^{5} \tilde{\gamma}^{l}\; \;
J^{a}(x) = \nonumber
\\
= +\; \mbox{Sp} \;  \; \left [ \gamma^{5} \;  \gamma^{l}\; \left (
\;
 + i \; A ^{*} -  \gamma ^{5} \; B^{*}  +   \gamma^{p} \; A^{*} _{p}  +
           i \; \gamma ^{5} \gamma ^{p}  \; B_{p}^{*}
+           i \; \sigma^{bc} \;  F ^{*}_{bc}   \; \right )\;
\times \right.
\nonumber
\\
\left. \times \gamma^{a}\; \left (\;  - i \; A  +  \gamma ^{5} \;
B  +   \gamma^{s} \; A _{s}  +
           i \; \gamma ^{s} \gamma ^{5} \; B_{s}
+ i \; \sigma^{mn}\;  F _{mn}   \; \right ) \; \right ] \; ,
\label{3.4d}
\end{eqnarray}
\begin{eqnarray}
L^{(a)kl}(x) = \mbox{Sp} \; \tilde{\sigma}^{kl} \; J^{a}(x) =
\nonumber
\\
= +\; \mbox{Sp} \;  \; \left [  \sigma^{kl}\; \left ( \;
 + i \; A ^{*} -  \gamma ^{5} \; B^{*}  +   \gamma^{p} \; A^{*} _{p}  +
           i \; \gamma ^{5} \gamma ^{p}  \; B_{p}^{*}
+           i \; \sigma^{bc} \;  F ^{*}_{bc}   \; \right )\;
\times \right.
\nonumber
\\
\left. \times \gamma^{a}\; \left (\;  - i \; A  +  \gamma ^{5} \;
B  +   \gamma^{s} \; A _{s}  +
           i \; \gamma ^{s} \gamma ^{5} \; B_{s}
+ i \; \sigma^{mn}\;  F _{mn}   \; \right ) \; \right ] \; .
\label{3.4e}
\end{eqnarray}

The relationships  presuppose rather unwieldy
calculation. For us in the first place the most interesting are
ordinary particles with one  value of spin, 0 or 1,  and certain
intrinsic parity. So, instead of (3,4) we  will get the four
expansions  for more simple cases $S=0,\tilde{0},1,\tilde{1}$.

\underline{Particle $S=0$}
\begin{eqnarray}
\left. \begin{array}{r} j^{(a)}(x) = -\; \mbox{Sp} \;  \; \left [
(\;
 + i \; A ^{*}  +   \gamma^{p} \; A^{*} _{p}\;       )\;
\gamma^{a}\;
(\; - i \; A  +     \gamma^{s} \; A _{s} \; )  \; \right ] \; , \\[3mm]
\nu^{(a)}(x) = -\; \mbox{Sp} \;  \; \left [  \gamma^{5} \; ( \; +
i \; A ^{*}  +   \gamma^{p} \; A^{*} _{p} \;) \; \gamma^{a}\; ( \;
- i \; A  +   \gamma^{s} \; A _{s} +
\; ) \; \right ]   \; ,\\[3mm]
T^{(a)l}(x) = +\; \mbox{Sp} \;  \; \left [  \gamma^{l}\;
 (\;  + i \; A ^{*}  +   \gamma^{p} \; A^{*} _{p}  \;  )\;
\gamma^{a}\;
 (\;  - i \; A    +   \gamma^{s} \; A _{s}   \; ) \; \right ] \; ,\\[3mm]
\nu^{(a)l}(x) = +\; \mbox{Sp} \;  \; \left [ \gamma^{5} \;
\gamma^{l}\; (
 + i \; A ^{*} +   \gamma^{p} \; A^{*} _{p}    )\;
\gamma^{a}\;
(\;  - i \; A   +   \gamma^{s} \; A _{s} \; ) \; \right ] \; ,\\[3mm]
L^{(a)kl}(x) = +\; \mbox{Sp} \;  \; \left [  \sigma^{kl}\;  (\;
 + i \; A ^{*}  +   \gamma^{p} \; A^{*} _{p}  \;  )\;
\gamma^{a}\;
 (\;  - i \; A   +   \gamma^{s} \; A _{s} \; ) \; \right ] \; .
\end{array} \right.
\label{3.5}
\end{eqnarray}

 \underline{Particle $S=\tilde{0}$}
\begin{eqnarray}
\left. \begin{array}{r} j^{(a)}(x) =  - \; \mbox{Sp} \;  \; \left
[  ( \;
 -  \gamma ^{5} \; B^{*}  +   i \; \gamma ^{5} \gamma ^{p}  \; B_{p}^{*}
   \;  )\; \gamma^{a}\;(\; +\gamma ^{5} \; B + i\;\gamma ^{s} \gamma ^{5} \; B_{s}
 \; ) \; \right ] \; , \\ [3mm]
\nu^{(a)}(x) = -\; \mbox{Sp} \;  \; \left [  \gamma^{5} \; ( \;
 -  \gamma ^{5} \; B^{*} + i \; \gamma ^{5} \gamma ^{p}  \; B_{p}^{*}
\; )\; \gamma^{a}\; (\;   +  \gamma ^{5} \; B  + i \; \gamma ^{s}
\gamma ^{5} \; B_{s}
  \; ) \; \right ] \; ,\\[3mm]
T^{(a)l}(x) = +\; \mbox{Sp} \;  \; \left [  \gamma^{l}\; ( \;
 -  \gamma ^{5} \; B^{*}  +  i \; \gamma ^{5} \gamma ^{p}  \; B_{p}^{*}
   \;  )\;  \gamma^{a}\;
 (\;  +  \gamma ^{5} \; B +  i \; \gamma ^{s} \gamma ^{5} \; B_{s}
\;  ) \; \right ] \; ,\\[3mm]
\nu^{(a)l}(x) = +\; \mbox{Sp} \;  \; \left [ \gamma^{5} \;
\gamma^{l}\; ( \;
 -  \gamma ^{5} \; B^{*}  +
           i \; \gamma ^{5} \gamma ^{p}  \; B_{p}^{*}
 \;  )\;  \gamma^{a}\;
(\;  +  \gamma ^{5} \; B  +   i \; \gamma ^{s} \gamma ^{5} \;
B_{s}
  \;  ) \; \right ] \; , \\[3mm]
L^{(a)kl}(x) = +\; \mbox{Sp} \;  \; \left [  \sigma^{kl}\;  ( \;
 -  \gamma ^{5} \; B^{*}  +  i \; \gamma ^{5} \gamma ^{p}  \; B_{p}^{*}
   \; )\;  \gamma^{a}\;
(\;  +  \gamma ^{5} \; B  +
           i \; \gamma ^{s} \gamma ^{5} \; B_{s}
   \;  ) \; \right ] \; .
\end{array} \right.
\label{3.6}
\end{eqnarray}

\underline{Particle $S=1$}
\begin{eqnarray}
\left. \begin{array}{r} j^{(a)}(x) = -\; \mbox{Sp} \;  \; \left [
( \;
  +   \gamma^{p} \; A^{*} _{p} + i\; \sigma^{bc} \;  F ^{*}_{bc}\; )\;
 \gamma^{a}\;
(\;   +   \gamma^{s} \; A _{s} + i \; \sigma^{mn}\;  F _{mn}   \; ) \; \right ] \; ,\\[3mm]
\nu^{(a)}(x) = -\; \mbox{Sp} \;  \; \left [  \gamma^{5} \; ( \; +
\gamma^{p} \; A^{*} _{p}  + i \; \sigma^{bc} \;  F ^{*}_{bc}   \;
)\; \gamma^{a}\;
(\;   +   \gamma^{s} \; A _{s}  + i \; \sigma^{mn}\;  F _{mn}   \; ) \; \right ] \; ,\\[3mm]
T^{(a)l}(x) = +\; \mbox{Sp} \;  \; \left [  \gamma^{l}\; ( \;
+\gamma^{p} \; A^{*} _{p}  + i \; \sigma^{bc} \;  F ^{*}_{bc}   \;
)\;
 \gamma^{a}\;
( \; + \gamma^{s} \; A _{s} + i \; \sigma^{mn}\;  F _{mn}   \; ) \; \right ] \; ,\\[3mm]
\nu^{(a)l}(x) = +\; \mbox{Sp} \;  \; \left [ \gamma^{5} \;
\gamma^{l}\; ( \; +   \gamma^{p} \; A^{*} _{p}  + i\;\sigma^{bc}
\;  F ^{*}_{bc}   \; )\; \gamma^{a}\;
(\;  + \gamma^{s} \; A _{s} + i \; \sigma^{mn}\;  F _{mn}   \; ) \; \right ] \;,\\[3mm]
L^{(a)kl}(x) = +\; \mbox{Sp} \;  \; \left [  \sigma^{kl}\;  ( \; +
\gamma^{p} \; A^{*} _{p} + i \; \sigma^{bc} \;  F ^{*}_{bc}   \;
)\; \gamma^{a}\; (\;  + \gamma^{s} \; A _{s}  + i \; \sigma^{mn}\;
F _{mn}   \;  ) \; \right ] \; .
\end{array} \right.
\label{3.7}
\end{eqnarray}

\underline{Particle  $S=\tilde{1}$}
\begin{eqnarray}
\left. \begin{array}{r} j^{(a)}(x) = -\; \mbox{Sp} \;  \; \left [
( \;
 + i \; \gamma ^{5} \gamma ^{p}  \; B_{p}^{*}
+  i \; \sigma^{bc} \;  F ^{*}_{bc}   \; )\;
 \gamma^{a}\;
(\; + i \; \gamma ^{s} \gamma ^{5} \; B_{s}
    + i \; \sigma^{mn}\;  F _{mn}   \;  ) \; \right ] \; ,\\[3mm]
\nu^{(a)}(x) = -\; \mbox{Sp} \;  \; \left  [  \gamma^{5} \; ( \;
 +  i \; \gamma ^{5} \gamma ^{p}  \; B_{p}^{*}
 +  i \; \sigma^{bc} \;  F ^{*}_{bc}   \; ) \;
\gamma^{a}\; (\;   + i \; \gamma ^{s} \gamma ^{5} \; B_{s}
      + i \; \sigma^{mn}\;  F _{mn}   \; ) \; \right ] \; ,\\[3mm]
T^{(a)l}(x) = +\; \mbox{Sp} \;  \; \left [  \gamma^{l}\;  ( \;
  + i \; \gamma ^{5} \gamma ^{p}  \; B_{p}^{*}
  + i \; \sigma^{bc} \;  F ^{*}_{bc}   \; )\;
 \gamma^{a}\;
(\;    + i \; \gamma ^{s} \gamma ^{5} \; B_{s}
       + i \; \sigma^{mn}\;  F _{mn}   \; ) \; \right ] \; ,\\[3mm]
\nu^{(a)l}(x) = +\; \mbox{Sp} \;  \; \left [ \gamma^{5} \;
\gamma^{l}\; ( \; +  i \; \gamma ^{5} \gamma ^{p}  \; B_{p}^{*} +
i \; \sigma^{bc} \;  F ^{*}_{bc}   \;  )\; \gamma^{a}\; (\;  + i
\; \gamma ^{s} \gamma ^{5} \; B_{s}
     + i \; \sigma^{mn}\;  F _{mn}   \;  ) \; \right ] \; ,\\[3mm]
L^{(a)kl}(x) = +\; \mbox{Sp} \;  \; \left [  \sigma^{kl}\;  ( \; +
i \; \gamma ^{5} \gamma ^{p}  \; B_{p}^{*} +   i \; \sigma^{bc} \;
F ^{*}_{bc}   \; )\;
 \gamma^{a}\;
(\;  + i \; \gamma ^{s} \gamma ^{5} \; B_{s}
     + i \; \sigma^{mn}\;  F _{mn}   \; ) \; \right ] \; .
\end{array} \right.
\label{3.8}
\end{eqnarray}

First, let us  consider the case of scalar particle $S=0$,
starting from the current $j^{(a)}(x)$ (take notice on factor
$1/4$)
\begin{eqnarray}
S=0: \qquad \mbox{Ток} \qquad j^{(a)}(x) = -\;{1 \over 4}\;
\mbox{Sp} \;  \; \left [  (\;
 + i \; A ^{*}  +   \gamma^{p} \; A^{*} _{p}\;       )\;
\gamma^{a}\; (\; - i \; A  +     \gamma^{s} \; A _{s} \; )  \;
\right ] \; ,
\label{3.9a}
\end{eqnarray}

\noindent an auxiliary table (the all non-zero traces of  matrix
combinations)
\begin{eqnarray}
\left. \begin{array}{rcc}
  &   \qquad  \gamma^{a}      & \qquad \gamma^{a} \gamma^{s} \\
I            &   \qquad     0            & \qquad  \mbox{Sp}\; (\gamma^{a} \gamma^{s}) \\
\gamma^{p}   &   \qquad \mbox{Sp}\;( \gamma^{p} \gamma^{a}) &
\qquad 0
\end{array} \right. \; ,
\nonumber
\end{eqnarray}

\noindent from this it follows
$$
S=0: \qquad j^{(a)}(x) = {1 \over i}  \; \left  ( \; A^{*} A^{a} -
A \; A^{a*}\;  \right ) \; . \eqno(3.9c)
$$

\noindent The current vanishes identically for a real-valued
scalar field. Now in the same manner consider the current
\begin{eqnarray}
S=0: \qquad   \nu^{(a)}(x) = -\;{1 \over 4} \; \mbox{Sp} \;  \;
\left [  \gamma^{5} \; ( \; + i \; A ^{*}  + \gamma^{p} \; A^{*}
_{p} \;) \; \gamma^{a}\; ( \; - i \; A  + \gamma^{s} \; A _{s}  +
\; ) \; \right ] \; ;
\label{3.10a}
\end{eqnarray}

\noindent an auxiliary table
\begin{eqnarray}
\left. \begin{array}{rcc}
  &   \qquad  \gamma^{a}      & \qquad \gamma^{a} \gamma^{s} \\
\gamma^{5}            &   \qquad     0            & \qquad  0 \\
\gamma^{5} \gamma^{p}   &   \qquad   0  & \qquad 0
\end{array} \right. \; ,
\nonumber
\end{eqnarray}

\noindent so that
\begin{eqnarray}
S=0: \qquad  \nu^{(a)}(x) \equiv    0 \; .
\label{3.10c}
\end{eqnarray}

Consider the current
\begin{eqnarray}
S=0: \qquad  T^{(a)l}(x) = +\; {1 \over 4} \; \mbox{Sp} \;  \;
\left [  \gamma^{l}\;
 (\;  + i \; A ^{*}  +   \gamma^{p} \; A^{*} _{p}  \;  )\;
\gamma^{a}\;
 (\;  - i \; A    +   \gamma^{s} \; A _{s}   \; ) \; \right ] \; ;
\label{3.11a}
\end{eqnarray}

\noindent an auxiliary table
\begin{eqnarray}
\left. \begin{array}{rcc}
  &   \qquad  \gamma^{a}      & \qquad \gamma^{a} \gamma^{s} \\
\gamma^{l}            &   \qquad     \mbox{Sp}\;  (\gamma^{l}\gamma^{a} )  & \qquad  0 \\
\gamma^{l} \gamma^{p}   &   \qquad   0  & \qquad \mbox{Sp}\;
(\gamma^{l}\gamma^{p}\gamma^{a} \gamma^{s})
\end{array} \right. \; ,
\nonumber
\end{eqnarray}

\noindent non-zero traces
\begin{eqnarray}
{1 \over 4} \;\mbox{Sp}\; (\gamma^{l}\gamma^{a}) = g^{la}\;,
\qquad {1 \over 4} \;\mbox{Sp}\;
(\gamma^{l}\gamma^{p}\gamma^{a}\gamma^{s}) = g^{lp} g^{as} -
g^{la}g^{ps} + g^{ls} g^{pa} \; ,
\nonumber
\end{eqnarray}

\noindent the  conserved current is
\begin{eqnarray}
S=0: \qquad  T^{(a)l}(x) = A^{*} A\; g^{al}  - g^{al}\;(
A^{*}_{s}\; A^{s})  + ( A^{a*} \; A^{l}  + A^{a}\; A^{l*} ) \; .
\label{3.11c}
\end{eqnarray}

\noindent This quantity differs from zero both for complex- and
real-valued scalar field; it is symmetrical with respect to
indices  $al$, and evidently should  be considered as the
energy-momentum tensor for a scalar  particle. For the current
\begin{eqnarray}
S=0: \qquad  \nu^{(a)l}(x) = +\; \mbox{Sp} \; \; \left [
\gamma^{5} \;  \gamma^{l}\; (
 + i \; A ^{*} +   \gamma^{p} \; A^{*} _{p}    )\;
\gamma^{a}\; (\;  - i \; A   +   \gamma^{s} \; A _{s} \; ) \;
\right ]   \; ,
\label{3.12a}
\end{eqnarray}

\noindent an auxiliary table
\begin{eqnarray}
\left. \begin{array}{rcc}
  &   \qquad  \gamma^{a}  & \qquad \gamma^{a} \gamma^{s} \\
\gamma^{5}\gamma^{l}      & \qquad  0    & \qquad  0 \\
\gamma^{5} \gamma^{l} \gamma^{p}   &   \qquad   0  & \qquad
\mbox{Sp}\; (\gamma^{5} \gamma^{l}\gamma^{p}\gamma^{a}\gamma^{s})
\end{array} \right. \; ,
\nonumber
\end{eqnarray}

\noindent and the current expression is
\begin{eqnarray}
S=0: \qquad \nu^{(a)l}(x) = i\; \epsilon^{alps} \; A^{*}_{p}\;
A_{s} \; .
\label{3.12c}
\end{eqnarray}

\noindent The  current vanishes for all real-valued  fields. And
now  consider the last current
\begin{eqnarray}
S=0: \qquad   L^{(a)kl}(x) = +\; \mbox{Sp} \;  \; \left [
\sigma^{kl}\;  (\;
 + i \; A ^{*}  +   \gamma^{p} \; A^{*} _{p}  \;  )\;
\gamma^{a}\;
 (\;  - i \; A   +   \gamma^{s} \; A _{s} \; ) \; \right ] \; ,
\label{3.13a}
\end{eqnarray}

\noindent an auxiliary table
\begin{eqnarray}
\left. \begin{array}{rcc}
  &   \qquad  \gamma^{a}  & \qquad \gamma^{a} \gamma^{s} \\
\sigma^{kl}       & \qquad  0    & \qquad  \mbox{Sp} \;(
\sigma^{kl} \gamma^{a} \gamma^{s})
\\
\sigma^{kl} \gamma^{p}   &   \qquad   \mbox{Sp} \; ( \sigma^{kl}
\gamma^{p} \gamma^{a} ) & \qquad 0
\end{array} \right. \; ,
\label{3.13b}
\end{eqnarray}

\noindent the non-zero traces
\begin{eqnarray}
{1 \over 4} \; \mbox{Sp} \;(\sigma^{kl} \gamma^{a} \gamma^{s}) =
{1\over 2} \; ( -g^{ka} g^{ls} + g^{ks} g^{la} ) \; ,
\qquad
  {1
\over 4} \; \mbox{Sp} \; (\sigma^{kl} \gamma^{s} \gamma^{a}) =
{1\over 2} \; ( -g^{ks} g^{la} + g^{ka} g^{ls} ) \; ,
\nonumber
\end{eqnarray}

\noindent an expression for the current is
\begin{eqnarray}
S=0: \qquad   L^{(a)kl}(x) = {i \over 2} \;
 \left [ \; (A^{*}\;A^{k} + A\; A^{k*} )\; g^{la} -
(A^{*}\;A^{l} + A\; A^{l*} )\; g^{ka}  \; \right ]\; .
\label{3.13c}
\end{eqnarray}

\noindent The current is non-zero both for complex- and
real-valued field, it should be called the angular moment.
Collecting together all results (take notice that  we have no
interpretation for the current $\nu^{(a)l}(x)$):

\underline{$S=0:$}
\begin{eqnarray}
\left. \begin{array}{l}
j^{(a)}(x) = - i  \; \left  ( \; A^{*} A^{a} - A \; A^{a*}\;  \right ) \; , \\[2mm]
 \nu^{(a)}(x) \equiv    0 \; , \\[2mm]
T^{(a)l}(x) = A^{*} A \; g^{al} - g^{al}\;( A^{*}_{s}\; A^{s})  +
( A^{a*} \; A^{l}
+ A^{a}\; A^{l*} ) \; , \\[2mm]
\nu^{(a)l}(x) = i\; \epsilon^{alps} \; A^{*}_{p}\; A_{s} \; , \\[2mm]
L^{(a)kl}(x) =
 (i/2) \;
 \left [ \; (A^{*}\;A^{k} + A\; A^{k*} )\; g^{la} -
(A^{*}\;A^{l} + A\; A^{l*} )\; g^{ka}  \; \right ]\; .
\end{array} \right.
\label{3.14}
\end{eqnarray}

The same for pseudo scalar particle will look (and again, we have
no interpretation for the current $\nu^{(a)l}(x)$)

 \underline{$S=\tilde{0}:$}
\begin{eqnarray}
\left. \begin{array}{l}
j^{(a)}(x) = + i  \; \left  ( \; B^{*} B^{a} - B \; B^{a*}\;  \right ) \; , \\[2mm]
 \nu^{(a)}(x) \equiv    0 \; , \\[2mm]
T^{(a)l}(x) = B^{*} B \; g^{al} - g^{al}\;( B^{*}_{s}\; B^{s})  +
( B^{a*} \; B^{l}
+ B^{a}\; B^{l*} ) \; , \\[2mm]
\nu^{(a)l}(x) = -i\; \epsilon^{alps} \; B^{*}_{p}\; B_{s} \; , \\[2mm]
L^{(a)kl}(x) =
 (i/2) \;
 \left [ \; (B^{*}\;B^{k} + B\; B^{k*} )\; g^{la} -
(B^{*}\;B^{l} + B\; B^{l*} )\; g^{ka}  \; \right ]\; .
\end{array} \right.
\label{3.15}
\end{eqnarray}

Now  let us consider the currents for a vector particle.
\begin{eqnarray}
S=1: \qquad j^{(a)}(x) = -{1 \over 4} \; \mbox{Sp} \;  \; \left [
( \;
  +   \gamma^{p} \; A^{*} _{p} + i\; \sigma^{bc} \;  F ^{*}_{bc}\; )\;
 \gamma^{a}\;
(\;   +   \gamma^{s} \; A _{s} + i \; \sigma^{mn}\;  F _{mn}   \;
) \; \right ] \;
\label{3.16a}
\end{eqnarray}

\noindent an auxiliary table looks as
\begin{eqnarray}
\left. \begin{array}{rcc}
&\qquad \gamma^{a} \gamma^{s} & \qquad \gamma^{a} \sigma^{mn}  \\
\gamma^{p} &  \qquad 0 & \qquad \mbox{Sp} \;(\gamma^{p} \gamma^{a} \sigma^{mn}) \\
\sigma^{bc} & \qquad \mbox{Sp} \; (\sigma^{bc} \gamma^{a}
\gamma^{s} ) & \qquad 0
\end{array} \right.
\nonumber
\end{eqnarray}

\noindent non-zero traces are
\begin{eqnarray}
{1\over 4} \; \mbox{Sp}\; (\gamma^{s} \gamma^{a} \sigma^{mn})= {1
\over 2}\;( -g^{ms}g^{na}+ g^{ma} g^{ns})  \; , \nonumber
\\
 {1\over
4} \; \mbox{Sp}\; (\gamma^{a} \gamma^{s} \sigma^{mn})= {1 \over
2}\;( -g^{ma}g^{ns}+ g^{ms} g^{na})  \; , \nonumber
\end{eqnarray}

\noindent so that
\begin{eqnarray}
S=1: \qquad j^{(a)} = +i \; (A^{*}_{n}F^{na} - A_{n}F^{na*} ) \;
\label{3.16c}
\end{eqnarray}

\noindent which  vanishes for a real-valued field.
For the current
\begin{eqnarray}
S=1: \qquad \nu^{(a)}(x) = -{1 \over 4} \; \mbox{Sp} \;  \; \left
[  \gamma^{5} ( \;
  +   \gamma^{p} \; A^{*} _{p} + i\; \sigma^{bc} \;  F ^{*}_{bc}\; )\;
 \gamma^{a}\;
(\;   +   \gamma^{s} \; A _{s} + i \; \sigma^{mn}\;  F _{mn}   \;
) \; \right ] \;  ,
\label{3.17a}
\end{eqnarray}

\noindent an auxiliary table
\begin{eqnarray}
\left. \begin{array}{rcc}
&\qquad \gamma^{a} \gamma^{s} & \qquad \gamma^{a} \sigma^{mn}  \\
\gamma^{5} \gamma^{p} &  \qquad 0 & \qquad
\mbox{Sp} \;(\gamma^{5} \gamma^{p} \gamma^{a} \sigma^{mn}) \\
\gamma^{5} \sigma^{bc} & \qquad \mbox{Sp} \; ( \gamma^{5}
\sigma^{bc} \gamma^{a} \gamma^{s} ) & \qquad 0
\end{array} \right. ,
\nonumber
\end{eqnarray}

\noindent non-zero traces
\begin{eqnarray}
{1\over 4} \; \mbox{Sp}\; (\gamma^{5}  \gamma^{p} \gamma^{a}
\sigma^{mn})= {i \over 2}\; \epsilon^{pamn}  \; , \qquad \mbox{Sp}
\; ( \gamma^{5} \sigma^{bc} \gamma^{a} \gamma^{s} ) = {i \over 2}
\;\epsilon^{asbc} \; ,
\nonumber
\end{eqnarray}

\noindent so that the current is
\begin{eqnarray}
S=1: \qquad \nu^{(a)} = -{1 \over 2}  \; (A^{*}_{s}F_{mn} -
A_{s}F^{*}_{mn} ) \; \epsilon^{asmn}\;  .
\label{3.17c}
\end{eqnarray}

\noindent It differs from zero only for complex-valued fields.
Now for
\begin{eqnarray}
S=1: \qquad T^{(a)l}(x) = +{1 \over 4}\;  \mbox{Sp} \;   \left
[  \gamma^{l}\; (  +\gamma^{p}  A^{*} _{p}  + i  \sigma^{bc}
  F ^{*}_{bc}   \; )\;
 \gamma^{a}\;
(  + \gamma^{s}  A _{s} + i  \sigma^{mn}  F _{mn}    )
\; \right ] \; ,
\label{3.18a}
\end{eqnarray}

\noindent an auxiliary table looks
\begin{eqnarray}
\left. \begin{array}{rcc}
&\qquad \gamma^{a} \gamma^{s} & \qquad \gamma^{a} \sigma^{mn}  \\
\gamma^{l} \gamma^{p} &  \qquad  \mbox{Sp}\; ( \gamma^{l}
\gamma^{p} \gamma^{a} \gamma^{s} )
 & \qquad 0 \\
\gamma^{l} \sigma^{bc} & \qquad  0 & \qquad \mbox{Sp}
\;(\gamma^{l} \sigma^{bc}\gamma^{a} \sigma^{mn})
\end{array} \right. \; .
\nonumber
\end{eqnarray}

\noindent In order to obtain the formula fox six-matrix product
let us write down the  formulas
\begin{eqnarray}
\gamma^{l} \sigma^{bc}  = {1 \over 2}\; ( - \gamma^{b} g^{lc} +
\gamma^{c} g^{lb} + i\gamma^{5} \epsilon^{lbcd} \gamma_{d} ) \; ,
\nonumber
\\
\gamma^{a} \sigma^{mn} = {1 \over 2}\; ( - \gamma^{m} g^{an} +
\gamma^{n} g^{am} + i\gamma^{5} \epsilon^{amne} \gamma_{e} ) \; ,
\nonumber
\end{eqnarray}

\noindent and then
\begin{eqnarray}
(\gamma^{l} \sigma^{bc})(\gamma^{a} \sigma^{mn}) =
 {1 \over 4} ( - \gamma^{b} g^{lc} + \gamma^{c} g^{lb} +
i\gamma^{5} \epsilon^{lbcd} \gamma_{d} )  ( - \gamma^{m} g^{an}
+ \gamma^{n} g^{am} + i\gamma^{5} \epsilon^{amne} \gamma_{e} )
=
\nonumber
\\
{1 \over 4} \left [  -g^{lc} ( - \gamma^{b} \gamma^{m} g^{an}
+ \gamma^{b} \gamma^{n} g^{am} + i \gamma^{b} \gamma^{5}
\epsilon^{amne} \gamma_{e} ) + g^{lb}( - \gamma^{c} \gamma^{m} g^{an} + \gamma^{c}
\gamma^{n} g^{am} + i \gamma^{c} \gamma^{5} \epsilon^{amne}
\gamma_{e} ) + \right.
\nonumber
\\
\left. +  i \epsilon^{lbcd} \; ( - \gamma^{5} \gamma_{d}
\gamma^{m} g^{an} + \gamma^{5} \gamma_{d} \gamma^{n} g^{am} + i
\gamma^{5} \gamma_{d} \gamma^{5} \epsilon^{amne} \gamma_{e} )
\right ]  .
\nonumber
\end{eqnarray}

\noindent Taking the trace of such a matrix expression we will
arrive at the formula we need
\begin{eqnarray}
{1\over 4} \; \mbox{Sp}  \; \left [ (\gamma^{l}
\sigma^{bc})(\gamma^{a} \sigma^{mn}) \; \right ]=
\nonumber
\\
= {1 \over 4}\; \left [  -g^{lc} ( - g^{bm} g^{an} + g^{bn} g^{am}
)  +
 g^{lb} ( - g^{cm}  g^{an} + g^{cn}  g^{am} )  +
 \epsilon^{lbcd} \;   \epsilon^{amne} g_{de}   \right ] \; .
\label{3.19b}
\end{eqnarray}

\noindent Now from (\ref{3.18a}) we get for the current
\begin{eqnarray}
T^{(a)l} = A^{*}_{p}A_{s} \; {1 \over 4}\; \mbox{Sp} \; (
\gamma^{l} \gamma^{p} \gamma^{a} \gamma^{s} ) - F_{bc}^{*} F_{mn}
\; {1 \over 4}\; \mbox{Sp}  \; \left [ (\gamma^{l}
\sigma^{bc})(\gamma^{a} \sigma^{mn}) \; \right ] =
\nonumber
\\
= A^{*}_{p}A_{s} \; ( g^{lp} g^{as} - g^{la} g^{ps} + g^{ls}
g^{pa}) -
\nonumber
\\
- {1 \over 4}\; F_{bc}^{*} F_{mn} \; \left [  -g^{lc} ( - g^{bm}
g^{an} + g^{bn}  g^{am}  )  +
 g^{lb} ( - g^{cm}  g^{an} + g^{cn}  g^{am} )  +
 \epsilon^{lbcd} \;   \epsilon^{amne} g_{de}   \right ]
\nonumber
\end{eqnarray}

\noindent  and further
\begin{eqnarray}
T^{(a)l} = -g^{al} \; A^{*}_{p}A^{p} + ( A^{a*}A^{l} + A^{a}A^{l*}
) +
 F^{*ln} F_{n}^{\;\;\;a} - {1 \over 4}\; F_{bc}^{*} F_{mn} \;
 \epsilon^{lbcd} \;   \epsilon^{amne} g_{de} \; .
\label{3.20a}
\end{eqnarray}

\noindent Here the  last term may be changed to
\begin{eqnarray}
- {1 \over 4}\; F_{bc}^{*} F_{mn} \;
 \epsilon^{lbcd} \;   \epsilon^{amne} g_{de} =
+ {1 \over 4} \; F_{bc}^{*} F_{mn} \; \left | \begin{array}{ccc}
g^{la} & g^{lm} & g^{ln} \\
g^{ba} & g^{bm} & g^{bn} \\
g^{ca} & g^{cm} & g^{cn}
\end{array} \right | =
 {1 \over 2}\; F^{*mn} F_{mn} +  F^{ln} F^{*\;a} _{n} \; .
\nonumber
\end{eqnarray}

\noindent With the help of that, from eq. (\ref{3.20a}) it follows
\begin{eqnarray}
T^{(a)l} = -g^{al} \; A^{*}_{p}A^{p} + ( A^{a*}A^{l} + A^{a}A^{l*}
) +
 {1 \over 2}\; F^{*mn} F_{mn} +( F^{*ln} F_{n}^{\;\;\;a}  +
F^{ln} F^{*\;a} _{n} ) \; .
\label{3.20c}
\end{eqnarray}

\noindent The tensor is symmetrical under indices  $al$, it
differs from zero for both complex- and real-valued fields, and
may be  considered as the energy-momentum tensor for a vector
$S=1$ particle.
Now consider another current
\begin{eqnarray}
S=1: \qquad \nu^{(a)l}(x) = +\; \mbox{Sp} \;  \; \left [
\gamma^{5}   \gamma^{l} (\; +   \gamma^{p}  A^{*} _{p}  +
i\;\sigma^{bc}   F ^{*}_{bc}   \; )\; \gamma^{a} (  + \gamma^{s} A
_{s} + i \; \sigma^{mn}  F _{mn}   \; ) \; \right ] \; ,
\label{3.21a}
\end{eqnarray}

\noindent  an auxiliary table is
\begin{eqnarray}
\left. \begin{array}{rcc}
&\qquad \gamma^{a} \gamma^{s} & \qquad \gamma^{a} \sigma^{mn}  \\
\gamma^{5} \gamma^{l} \gamma^{p} &  \qquad  \mbox{Sp}\; (
\gamma^{5} \gamma^{l} \gamma^{p} \gamma^{a} \gamma^{s} )
 & \qquad 0 \\
\gamma^{5} \gamma^{l} \sigma^{bc} & \qquad  0 & \qquad \mbox{Sp}
\;(\gamma^{5} \gamma^{l} \sigma^{bc}\gamma^{a} \sigma^{mn})
\end{array} \right. \; .
\nonumber
\end{eqnarray}

\noindent We will need the trace-formula for second  term:
\begin{eqnarray}
{1 \over 4} \;\mbox{Sp}\; [ \gamma^{5} \;(\gamma^{l}
\sigma^{bc})(\gamma^{a} \sigma^{mn}) ] =
\nonumber
\\
=
 {1 \over 16}\;  \mbox{Sp}\; [ \gamma^{5} \; \left [ \; -g^{lc} (
- \gamma^{b} \gamma^{m} g^{an} + \gamma^{b} \gamma^{n} g^{am} + i
\gamma^{b} \gamma^{5} \epsilon^{amne} \gamma_{e} ) \; + \right.
\nonumber
\\
 +  g^{lb} ( - \gamma^{c} \gamma^{m} g^{an} + \gamma^{c}
\gamma^{n} g^{am} + i \gamma^{c} \gamma^{5} \epsilon^{amne}
\gamma_{e} )  +
\nonumber
\\
+
   i \epsilon^{lbcd}  ( - \gamma^{5} \gamma_{d}
\gamma^{m} g^{an} + \gamma^{5} \gamma_{d} \gamma^{n} g^{am} + i
\gamma^{5} \gamma_{d} \gamma^{5} \epsilon^{amne} \gamma_{e} ) \;
]  =
\nonumber
\\
= {i \over 4} \; \left  [ \;   (g^{lc}  \epsilon ^{amnb} -  g^{lb}
\epsilon ^{amnc} ) +
 \epsilon^{lbcd} \; (  \delta^{n}_{d} \; g^{am} - \delta^{m}_{d} \;g^{an} )  \; \right ] \; .
\nonumber
\end{eqnarray}

\noindent  with the help of that,  for the current we will find
\begin{eqnarray}
\nu^{(a)l}(x) = i\; A^{*}_{p}A_{s}  \; \epsilon^{alps} + {i\over
2} \; F_{bc}^{*} F_{mn} \; \left [\;
 g^{an}\;  \epsilon^{lbcm}  + g^{lb} \;  \epsilon ^{amnc}     \; \right ] \; .
\label{3.21d}
\end{eqnarray}

The conservation laws for the quantity $\nu^{(a)l}(x) $ may be
written down in detailed form:
\begin{eqnarray}
\left. \begin{array}{llr} \nu^{(a)0} : &\qquad \nu^{(0)0}  = &
 -i \; [\; + E^{*1} \;  B^{1}  +  E^{*2} \; B^{2} +
 E^{*3} \; B^{3}  - \mbox{c.c} \; ]\; ,\\
& \qquad  \nu^{(1)0}   = & + i \; [ \;  - A_{2}^{*} \;A_{3}  +  (
E^{*}_{2}\;E_{3} +  B^{*}_{2}\;B_{3})  -
\mbox{c.c} \; ] \; ,\\
& \qquad\nu^{(2)0}  = & + i \; [ \;  - A_{3}^{*} \;A_{1}  + (
E^{*}_{3}\; E_{1} +  B^{*}_{3} \; B_{1})  -
\mbox{c.c} \; ] \; ,\\
& \qquad \nu^{(3)0}  = & + i \; [ \;  - A_{1}^{*} \;A_{2}  +  (
E^{*}_{1}\; E_{2} +  B^{*}_{1} \; B_{2})  - \mbox{c.c}  \; ] \; ,
\end{array} \right.
\label{3.22a}
\end{eqnarray}
\begin{eqnarray}
\left. \begin{array}{llr} \nu^{(a)1} : &\qquad \nu^{(0)1} (x) = &
+ i \; [ \;  + A_{2}^{*} \;A_{3}  + ( E_{2}^{*}\; E_{3} +
B_{2}^{*} \; B_{3})  - \mbox{c.c}  \; ] \; , \\  &\qquad
\nu^{(1)1} (x) = & -i \; [ \;  - E^{*1} \; B^{1}  +  E^{*2}\;
B^{2} +
 E^{*3} \;
 B^{3}   - \mbox{c.c}\; ]  \; ,\\
&\qquad \nu^{(2)1} (x) = & i\; [\;  - \; A_{0}^{*} \; A_{3} + (
E_{2}^{*}  \;B_{1}  + E^{*}_{1} \; B_{2}   ) -
\mbox{c.c} \; ] \; ,\\
&\qquad \nu^{(3)1} (x) =& i\; [\; +\; A_{0}^{*} \; A_{2} + (
E^{*}_{3} \; B_{1}  + E_{1}^{*}  \;B_{3} ) - \mbox{c.c} \; ] \; ,
\end{array} \right.
\label{3.22b}
\end{eqnarray}
\begin{eqnarray}
\left. \begin{array}{llr} \nu^{(a)2} : &\qquad \nu^{(0)2} (x) =& +
i \; [ \;  + A_{3}^{*} \;A_{1}  + ( E^{*}_{3}\; E_{1} +  B^{*}_{3}
\; B_{1})  -
\mbox{c.c} \; ] \; , \\
&\qquad \nu^{(1)2} (x) =& i\; [\; +\; A_{0}^{*} \; A_{3} + (
E^{*}_{1} \; B_{2}  + E_{2}^{*}  \;B_{1} ) -
 \mbox{c.c} \; ] \; ,\\
&\qquad \nu^{(2)2} (x) =& -i \; [\; + E^{*1} \; B^{1}  -  E^{*2}
\; B^{2} +
 E^{*3} \;  B^{3}  - \mbox{c.c} \; ]   \; ,\\
&\qquad \nu^{(3)2} (x) =& i\; [\; - \; A_{0}^{*} \; A_{1} + (
E_{3}^{*}  \;B_{2} + E^{*}_{2}\;  B_{3}  ) -
  \mbox{c.c} \; ] \; ,
\end{array} \right.
\label{3.22c}
\end{eqnarray}
\begin{eqnarray}
\left. \begin{array}{llr} \nu^{(a)3} : &\qquad \nu^{(0)3} (x) =& +
i \; [ \;  + A_{1}^{*} \;A_{2}  + (  E^{*}_{1}\; E_{2} + B^{*}_{1}
\; B_{2})  -
\mbox{c.c} \; ] \; ,\\
&\qquad \nu^{(1)3} (x) =& i\; [\; -\; A_{0}^{*} \; A_{2} + (
E_{1}^{*}  \;B_{3} + E^{*}_{3} \; B_{1}   ) -
  \mbox{c.c} \; ] \; ,\\
&\qquad \nu^{(2)3} (x) =& i\; [\; +\; A_{0}^{*} \; A_{1} + (
E^{*}_{2} \; B_{3}  + E_{3}^{*}  \;B_{2} )
-   \mbox{c.c} \; ] \; ,\\
&\qquad \nu^{(3)3} (x) =& -i \; [\; + E^{*1} \; B^{1}  +  E^{*2}
\; B^{2} -
 E^{*3}\;  B^{3}  - \mbox{к.с} \; ]   \; .
\end{array} \right.
\label{3.22d}
\end{eqnarray}

And finally, let us obtain an explicit form of the current
\begin{eqnarray}
S=1: \qquad L^{(a)kl}(x) = + {1 \over 4} \; \mbox{Sp} ;  \; \left
[  \sigma^{kl}  (  +   \gamma^{p}  A^{*} _{p} + i
\sigma^{bc}   F ^{*}_{bc}     ) \gamma^{a} (  +
\gamma^{s}  A _{s}  + i  \sigma^{mn}  F _{mn}     ) \;
\right ] \; ,
\label{3.23a}
\end{eqnarray}

\noindent
 an auxiliary table is
\begin{eqnarray}
\left. \begin{array}{rcc}
&\qquad \gamma^{a} \gamma^{s} & \qquad \gamma^{a} \sigma^{mn}  \\
\sigma^{kl} \gamma^{p} &  \qquad  0 &  \qquad \mbox{Sp}\;
(\sigma^{kl} \gamma^{p}
 \; \gamma^{a} \sigma^{mn} )
 \\   \sigma^{kl} \sigma^{bc} & \qquad  \mbox{Sp} \;(\sigma^{kl}
\sigma^{bc} \gamma^{a} \gamma^{s}
 ) & \qquad 0
\end{array} \right. \; ,
\nonumber
\end{eqnarray}

\noindent non-zero traces are
\begin{eqnarray}
{1 \over 4} \; \mbox{Sp} \; (\sigma^{kl} \sigma^{bc}  \gamma^{a}
\gamma^{s} ) = {1 \over 4}\; \mbox{Sp} \;( \sigma^{bc} \gamma^{a}
\gamma^{s} \sigma^{kl}   ) =
\nonumber
\\
= {1 \over 16}\; \mbox{Sp} \; \left [ (\gamma^{b} g^{ca} -
\gamma^{c} g^{ba} + i \gamma^{5} \; \epsilon^{bcad} \gamma_{d} )
(-\gamma^{k} g^{sl} + \gamma^{l} g^{sk} + i \gamma^{5}
\epsilon^{skle}\gamma_{e}) \right ]=
\nonumber
\\
={1 \over 4} \; \left [ g^{ca} (- g^{bk} g^{sl} + g^{bl} g^{sk} )
- g^{ba} ( - g^{ck} g^{sl} +  g^{cl} g^{sk} ) - \epsilon^{bcad}
\epsilon^{skle} g_{de} \right ]
\nonumber
\\
{1 \over 4 }  \; \mbox{Sp} \; (\sigma ^{kl} \gamma^{p}
 \gamma^{a} \sigma^{mn} ) =
\nonumber
\\
 = { 1 \over  16 } \;  \mbox{Sp} \;
[  ( \gamma^{k} g^{lp} - \gamma^{l} g^{kp} + i \gamma^{5}
\epsilon^{klpd} \gamma_{d} )
  (-\gamma^{m} g^{an} + \gamma^{n} g^{am} + i \gamma^{5} \epsilon^{amne} \gamma_{e} )
  ]=
\nonumber
\\
= {1 \over 4}\; \left [ g^{lp}( -g^{km} g^{an} + g^{kn} g^{am} ) -
g^{kp} ( - g^{lm} g^{an} +  g^{ln} g^{am} ) + \epsilon^{klpd}
\epsilon^{amne} g_{de})  \right ] \; .
\nonumber
\end{eqnarray}

\noindent With the help of two relations
\begin{eqnarray}
i \; F^{*}_{bc} A_{s} \; {1\over 4}\; \mbox{Sp} \; (\sigma^{kl}
\sigma^{bc}  \gamma^{a} \gamma^{s} ) =
\nonumber
\\
= {i\over 2} \left [ \; - A^{a} F^{*kl} + (F^{*ak} A^{l} -
F^{*al}A^{k}) + (- g^{ak} F^{*ls} A_{s} + g^{al} F^{*ks}A_{s} )
\right ] \; ,
\nonumber
\\
i A^{*}_{p} F_{mn}\; {1 \over 4 }  \; \mbox{Sp} \; (\sigma ^{kl}
\gamma^{p}
 \gamma^{a} \sigma^{mn} ) =
\nonumber
\\
{i \over 2} \; \left [ \; -A^{*a} F^{kl} + ( F^{ak} A^{*l} -
F^{al} A^{*k} ) + (-g^{ak} F^{ls} A^{*}_{s} + g^{al} F^{ks}
A^{*}_{s}) \right ]
\nonumber
\end{eqnarray}

\noindent one readily produces the expression for this current
\begin{eqnarray}
L^{(a)kl} = {i\over 2} \left \{ \; - ( A^{a} F^{*kl}  +  A^{*a}
F^{kl} ) +  [ \; (F^{*ak} A^{l} - F^{*al}A^{k}) +( F^{ak} A^{*l} -
F^{al} A^{*k} ) \; ]+ \right.
\nonumber
\\
\left.   + [ \;(- g^{ak} F^{*ls} A_{s} + g^{al} F^{*ks}A_{s} ) +
(-g^{ak} F^{ls} A^{*}_{s} + g^{al} F^{ks} A^{*}_{s}\; ] \;
 \right \} \; .
\label{3.24c}
\end{eqnarray}

Collecting together all results for a particle of $S=1$:
\begin{eqnarray}
\left. \begin{array}{ll} S=1:  & \qquad j^{(a)} = +i \; (A^{*}_{n}
\; F^{na} - A_{n} \;F^{na*} ) \; ,\\[3mm]
 & \qquad \nu^{(a)} =
-(1/2)  \; (A^{*}_{s}\; F_{mn} - A_{s} \; F^{*}_{mn} ) \;
\epsilon^{asmn}\;  , \\[3mm]
& \qquad T^{(a)l} = -g^{al} \; A^{*}_{p}\; A^{p} + ( A^{a*} \;
A^{l} +
A^{a} \; A^{l*} ) + \\[2mm]
& \qquad + (1 /2) \; F^{*mn} \; F_{mn} +( F^{*ln} \;
F_{n}^{\;\;\;a}  +
F^{ln} \;  F^{*\;a} _{n} ) \; , \\[3mm]
& \qquad \nu^{(a)l}(x) =
i\; A^{*}_{p}\; A_{s}  \; \epsilon^{alps} +  \\[2mm]
& \qquad +(i /2) \; F_{bc}^{*} \; F_{mn} \; (\;
 g^{an}\;  \epsilon^{lbcm}  + g^{lb} \;  \epsilon ^{amnc}     \; ) \; , \\[3mm]
 & \qquad
 L^{(a)kl} =
(i /2)  \left \{ \; - ( A^{a} \; F^{*kl}  +  A^{*a} \; F^{kl} ) +
\right. \\[2mm]
  & \qquad + [ \; (F^{*ak} \; A^{l} - F^{*al} \;
A^{k}) +( F^{ak}\; A^{*l} - F^{al} \;
A^{*k} ) \; ]+\\[2mm]
& \qquad \left.
 + [ \;(- g^{ak} \; F^{*ls} \; A_{s} + g^{al} \; F^{*ks} \; A_{s} ) +
(-g^{ak} \; F^{ls} \; A^{*}_{s} + g^{al} \; F^{ks} \;  A^{*}_{s}\;
] \;
 \right \} \; .
 \end{array} \right.
\end{eqnarray}

Take notice that we have no interpretation for two currents,
$\nu^{(a)} $  and $\nu^{(a)l}(x)$, which are non-zero  for
complex-valued  fields only. Here we will not consider in detail
the case of vector particle with another intrinsic parity,
particle of the type $S =\tilde{1}$, any additional  ideas will
not arise at this.

\end{document}